\documentclass[twocolumn,prl,aps,floatfix]{revtex4-1}

\usepackage{amsmath}
\usepackage{graphicx}
\usepackage{epstopdf}
\usepackage[usenames]{color}
\usepackage{xr}
\usepackage[breaklinks=true,
            pdfborder={0 0 1},
            colorlinks=true,
            linkcolor=black,
            citecolor=blue,
            urlcolor=blue]{hyperref}
\usepackage[normalem]{ulem}

\def\vp{v_\text{p}}
\def\Fp{F_\text{p}}
\def\Dr{D_\text{r}}
\def\fc{f_\text{c}}
\def\rhog{\rho_\text{g}}
\def\kt{k_\text{B}T}
\def\Pe{\text{Pe}}
\def\kin{k_\text{in}}
\def\kout{k_\text{out}}
\def\q6{q_\text{6}}

\begin{document}

\title{Structure and Dynamics of a Phase-Separating Active Colloidal
  Fluid}

\author{Gabriel S. Redner}
\author{Michael F. Hagan}
\email{hagan@brandeis.edu}
\author{Aparna Baskaran}
\email{aparna@brandeis.edu}

\affiliation{Martin Fisher School of Physics, Brandeis University,
  Waltham, MA, USA.}

\begin{abstract}
We examine a minimal model for an active colloidal fluid in the form
of self-propelled Brownian spheres that interact purely through
excluded volume with no aligning interaction. Using simulations and
analytic modeling, we quantify the phase diagram and separation
kinetics.  We show that this nonequilibrium active system undergoes an
analog of an equilibrium continuous phase transition, with a binodal
curve beneath which the system separates into dense and dilute phases
whose concentrations depend only on activity.  The dense phase is a
unique material that we call an active solid, which exhibits the
structural signatures of a crystalline solid near the crystal-hexatic
transition point, and anomalous dynamics including superdiffusive
motion on intermediate timescales.
\end{abstract}

\maketitle

\noindent Active fluids composed of self-propelled units occur in
nature on many scales ranging from cytoskeletal filaments and
bacterial suspensions to macroscopic entities such as insects, fish
and birds \cite{2010arXiv1010.5017V}.  These systems exhibit strange
and exciting phenomena such as dynamical self regulation
\cite{Gopinath2011}, clustering \cite{PhysRevE.74.030904}, anomalous
density fluctuations \cite{0295-5075-62-2-196}, unusual
rheological behavior \cite{PhysRevE.81.051908, PhysRevE.81.056307,
  PhysRevLett.101.068102}, and activity-dependent phase boundary
changes \cite{Shen2004}.  Motivated by these findings, recent
experiments have focused on realizing active fluids in nonliving
systems, using chemically propelled particles undergoing
self-diffusophoresis \cite{PhysRevLett.105.088304,
  doi:10.1021/ja047697z, PhysRevLett.99.178103}, Janus particles
undergoing thermophoresis \cite{PhysRevLett.105.268302, C1SM05960B},
as well as vibrated monolayers of granular particles
\cite{Narayan06072007, PhysRevLett.100.058001,
  PhysRevLett.105.098001}.
\begin{figure}[h!]
  \includegraphics[width=.49\columnwidth]{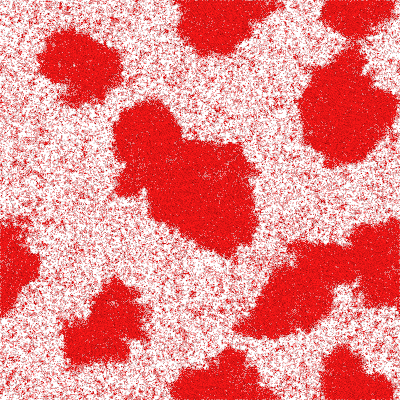}
  \includegraphics[width=.49\columnwidth]{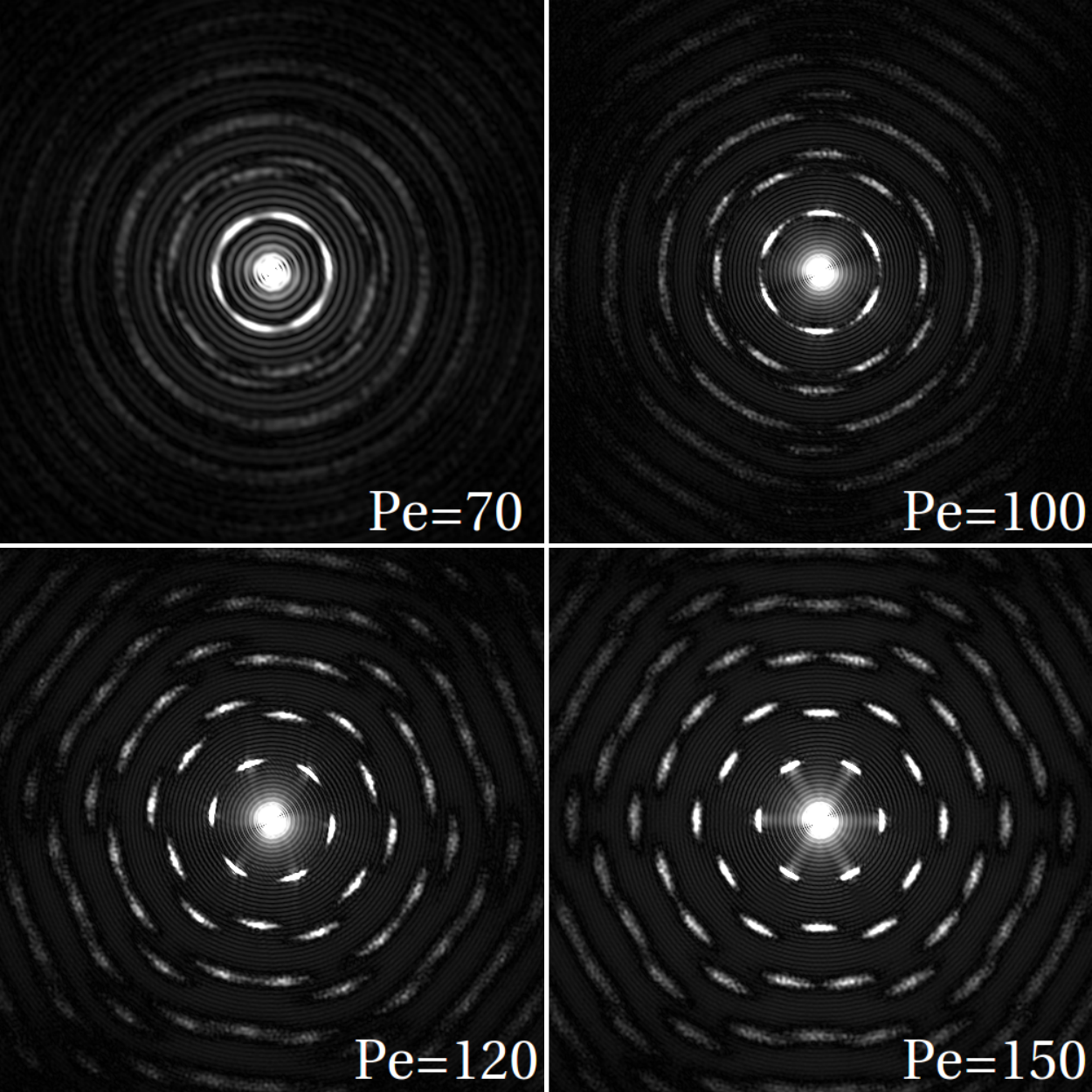} \\
  \includegraphics[width=.49\columnwidth]{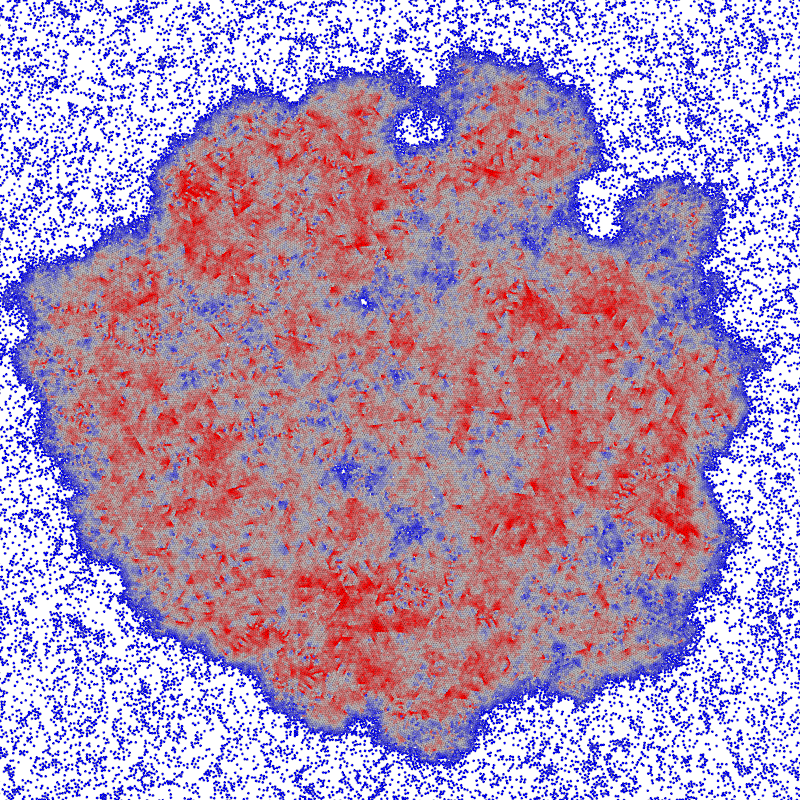}
  \includegraphics[width=.49\columnwidth]{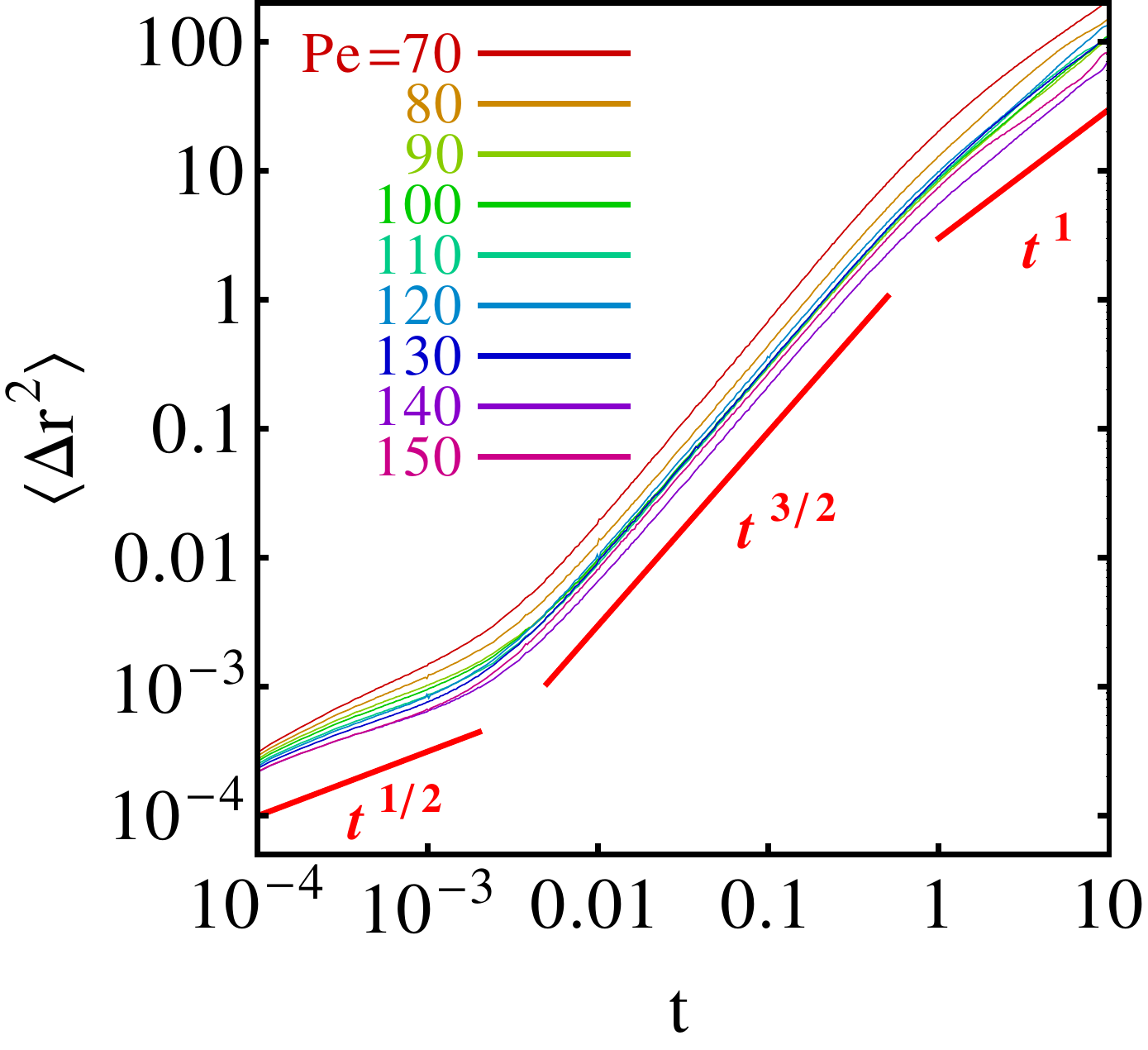}
  \caption{(color online) A visual summary of our results. Top left:
    Beyond critical density and activity levels the active colloidal
    fluid separates into dense and dilute phases.  The clusters
    coarsen over time (see \ref*{Supplement-fig:coarsening-movie} in
    \cite{SupplementalInformation}).  Top right: The static structure
    factor $S(\boldsymbol{k}) = \frac{1}{N} \left\langle \sum_{ij}
    e^{i \boldsymbol{k} \cdot \boldsymbol{r}_{ij}} \right\rangle$,
    restricted to the interiors of large clusters.  These signatures
    resemble those of a high temperature colloidal crystal near the
    crystal-hexatic phase transition. Bottom left: A heat map of the
    pressure in the active solid material. It is heterogeneous and
    highly dynamic, indicating that external stresses would produce a
    complex response.  Bottom right: Log-log plot of the mean square
    displacement of a tagged particle in the active solid.  At
    intermediate time scales, it exhibits anomalous superdiffusive
    transport.}
  \label{fig:summary}
\end{figure}

In this letter we explore a minimal active fluid model: a system of
self-propelled smooth spheres interacting by excluded volume alone and
confined to two dimensions.  Unlike self-propelled rods
\cite{PhysRevE.77.011920, PhysRevLett.101.268101,
  PhysRevLett.106.128101, PhysRevE.82.031904, C2SM06960A}, these
particles cannot interchange angular momentum and thus lack a mutual
alignment mechanism.  Recent simulation and experimental studies have
shown that this system exhibits giant number fluctuations
\cite{PhysRevLett.108.235702} and athermal phase separation
\cite{PhysRevLett.108.235702,2012arXiv1202.6264T} that are
characteristic of active fluids \cite{0295-5075-62-2-196,
  PhysRevLett.97.090602, Cates2010}. Here we employ extensive Brownian
dynamics simulations to characterize the phase diagram of this system
and we develop an analytic model that captures its essential
features. We show that this nonequilibrium system undergoes a
continuous phase transition, analogous to that of equilibrium systems
with attractive interactions, and that the phase separation kinetics
demonstrate equilibrium-like coarsening. These structural and dynamic
signatures of phase separation and coexistence enable an unequivocal
definition of phases in this nonequilibrium, active system. Finally,
we find that the dense phase is a dynamic new form of material that we
call an ``active solid''. This material exhibits structural properties
consistent with a 2D colloidal crystal near the crystal-hexatic
transition point \cite{PhysRevLett.104.205703, PhysRevE.77.041406},
but is characterized by such anomalous features as superdiffusive
transport at intermediate timescales and a heterogeneous and dynamic
stress distribution (see Fig.~(\ref{fig:summary})).

\noindent \emph{Model and Simulation Method}: Our system consists of
smooth spheres immersed in a solvent and confined to a plane, similar
to experimental systems of self-propelled colloids sedimented at an
interface \cite{2012arXiv1202.6264T}.  Each particle is self-propelled
with a constant force, and interactions between particles result from
isotropic excluded-volume repulsion only.  We include no mechanism for
explicit alignment or transmission of torques between particles.

The state of the system is represented by the positions and
self-propulsion directions $\{\boldsymbol{r}_i, \theta_i\}_{i=1}^N$ of
all particles.  Their evolution is governed by the coupled overdamped
Langevin equations:
\begin{align}
  \dot{\boldsymbol{r}}_i &= D \beta \left[
    \boldsymbol{F}_{\mathrm{ex}}(\{\boldsymbol{r}_i\}) + \Fp
    \hat{\boldsymbol{\nu}}_i \right] + 
  \sqrt{2 D} \, \boldsymbol{\eta}^T_i \\ 
  \dot{\theta}_i &= \sqrt{2 \Dr} \, \eta^R_i
\end{align}
Here $\boldsymbol{F}_{\mathrm{ex}}$ is an excluded-volume repulsive
force given by the WCA potential $V_{\mathrm{ex}} = 4 \epsilon \left[
  \left( \frac{\sigma}{r} \right)^{12} - \left( \frac{\sigma}{r}
  \right)^6 \right] + \epsilon$ if $r < 2^{\frac{1}{6}}$, and zero
otherwise \cite{weeks:5237}, with $\sigma$ the nominal particle
diameter.  We use $\epsilon = \kt$, but our results should be
insensitive to the exact strength and form of the potential.  $\Fp$ is
the magnitude of the self-propulsion force which in the absence of
interactions will move a particle with speed $\vp = D \beta \Fp$,
$\hat{\boldsymbol{\nu}}_i = (\cos \theta_i, \sin \theta_i)$, and
$\beta = \frac{1}{\kt}$.  $D$ and $\Dr$ are translational and
rotational diffusion constants, which in the low-Reynolds-number
regime are related by $\Dr = \frac{3D}{\sigma^2}$. The $\eta$ are
Gaussian white noise variables with $\left\langle \eta_i(t)
\right\rangle = 0$ and $\left\langle \eta_i(t) \eta_j(t')
\right\rangle = \delta_{ij} \delta(t-t')$.

We non-dimensionalized the equations of motion using $\sigma$ and
$\kt$ as basic units of length and energy, and $\tau =
\frac{\sigma^2}{D}$ as the unit of time.  Simulations employed the
stochastic Runge-Kutta method \cite{PhysRevE.60.2381} with maximum
timestep $2 \times 10^{-5} \tau$. Simulations mapping the phase
diagram were run with $15{,}000$ particles until time $100 \tau$,
while larger systems (up to $512{,}000$ particles) were used to
explore kinetics and material properties.  The simulation box was
square with periodic boundaries, with its size chosen to achieve the
desired density. The system is parametrized by two dimensionless
values, the packing fraction $\phi$ and the P\'eclet number, which in
our units is identical to the non-dimensionalized velocity ($\Pe = \vp
\frac{\tau}{\sigma}$). In this work, we varied $\phi$ from near-zero
to the hard-sphere close-packing value $\phi_{\mathrm{cp}} =
\frac{\pi}{2 \sqrt{3}}$, and $\Pe$ from zero to $150$.

\begin{figure}[tbp]
  \includegraphics[width=.51\columnwidth]{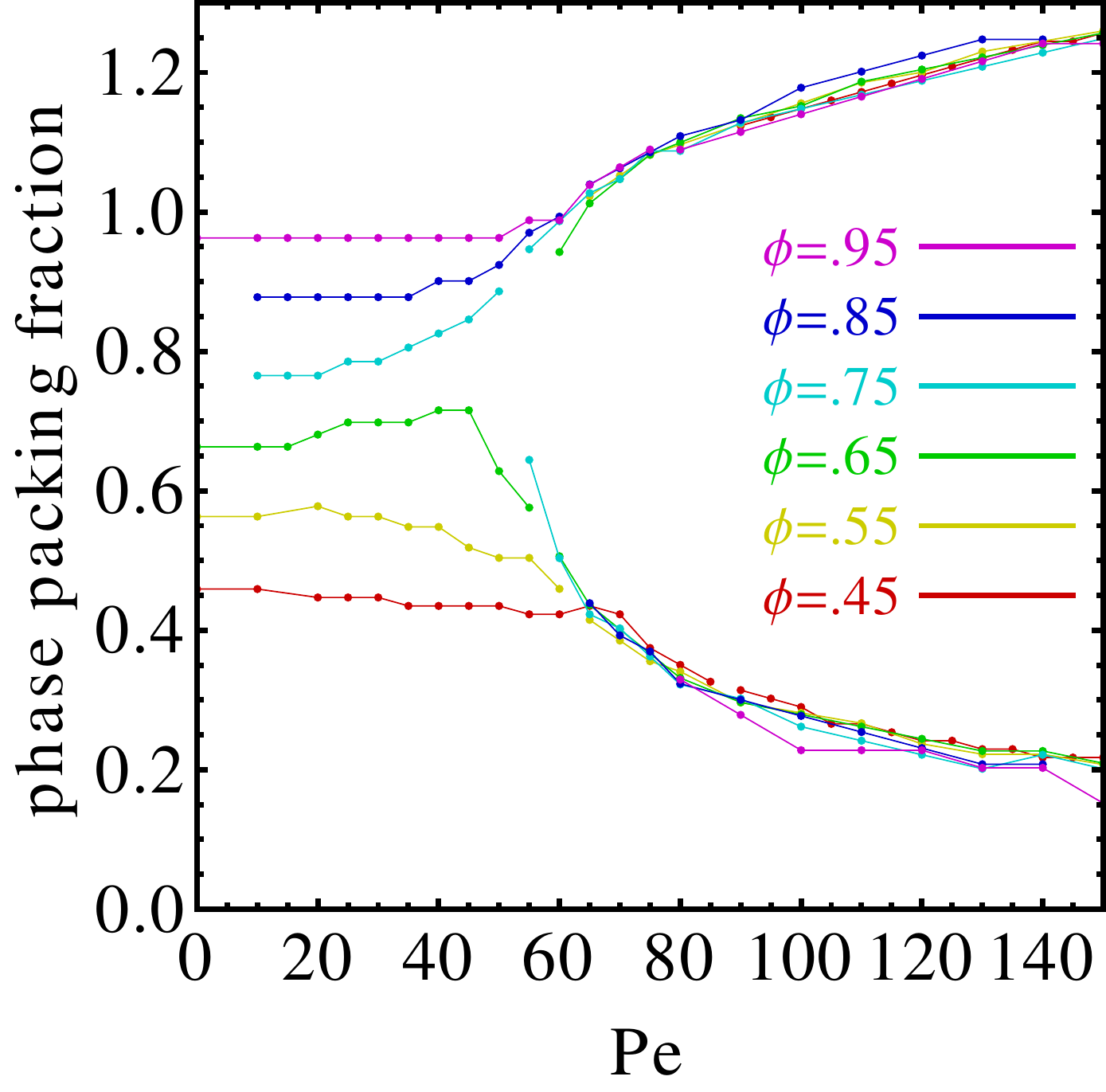}
  \raisebox{-1pt}{\includegraphics[width=.47\columnwidth]{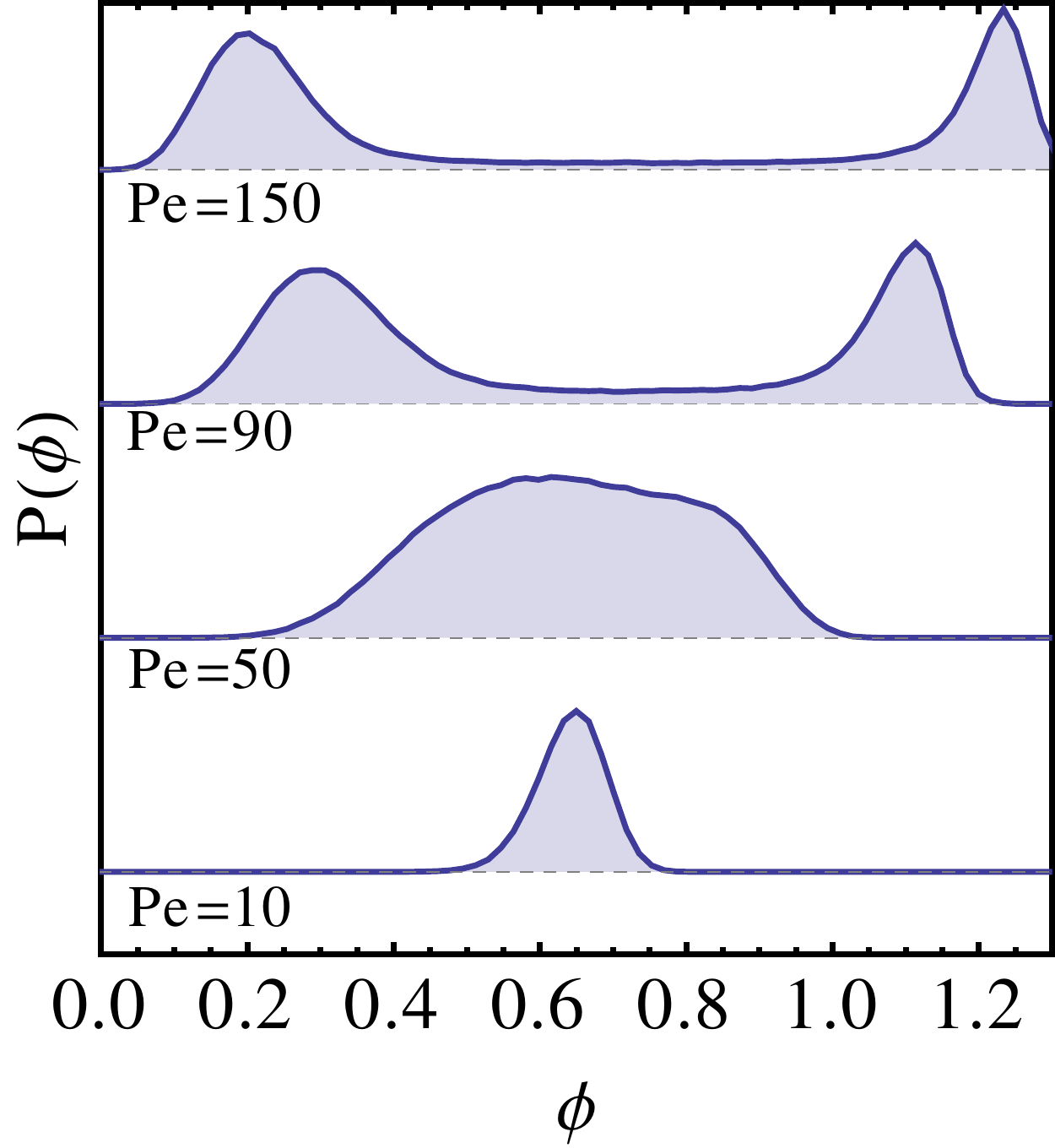}}
  \caption{(color online) Left: Phase densities as a function of
    P\'eclet number ($\Pe$) for a range of overall $\phi$.  At low
    $\Pe$ the system is single-phase, while at increased $\Pe$ it
    phase-separates.  The coexistence boundary is analogous to the
    binodal curve of an equilibrium fluid, with $\Pe$ acting as an
    attraction strength.  Right: Observed density distributions for
    various P\'eclet numbers.  In the single-phase region below $\Pe
    \approx 50$, $P(\phi)$ is peaked about the overall system density
    (here $\phi = 0.65$).  It broadens and flattens as the critical
    point is approached, and becomes bimodal as the system phase
    separates.}
  \label{fig:phase_separation}
\end{figure}

\noindent \emph{Phase Separation}: We first show that our results are
consistent with prior simulations \cite{PhysRevLett.108.235702} and
confirm that this system, despite the absence of aligning
interactions, shows the signature behaviors of an active fluid.  In
particular, the active spheres undergo nonequilibrium clustering
(Fig.~(\ref{fig:summary})) similar to other model active systems
\cite{Chate2008, PhysRevE.74.030904, C2SM06960A, PhysRevE.82.031904}.

We establish that this clustering is indeed athermal phase separation
by measuring the density in each phase at different parameter values
(Fig.~(\ref{fig:phase_separation}a)). We observe a binodal envelope
beyond which the system separates into two phases whose densities
collapse onto a single coexistence curve which is a function of
activity alone.  The phase diagram is thus analogous to that of an
equilibrium system of mutually attracting particles undergoing phase
separation, with $\Pe$ (playing the role of an attraction strength) as
the control parameter. This surprising result contradicts the
expectation that increased activity will destabilize aggregates and
suppress phase separation (as seen in \cite{Schwarz-Linek2012}) and
indicates that the effects of activity cannot be described by an
``effective temperature'' in this system.

Additionally, we identify a critical point at the apex of the bimodal
(near $\Pe=50$, $\phi=0.7$). In the vicinity of this point, the system
exhibits equilibrium-like critical phenomena which will be detailed in
a future publication.

\begin{figure}[tbp]
  \includegraphics[width=.49\linewidth]{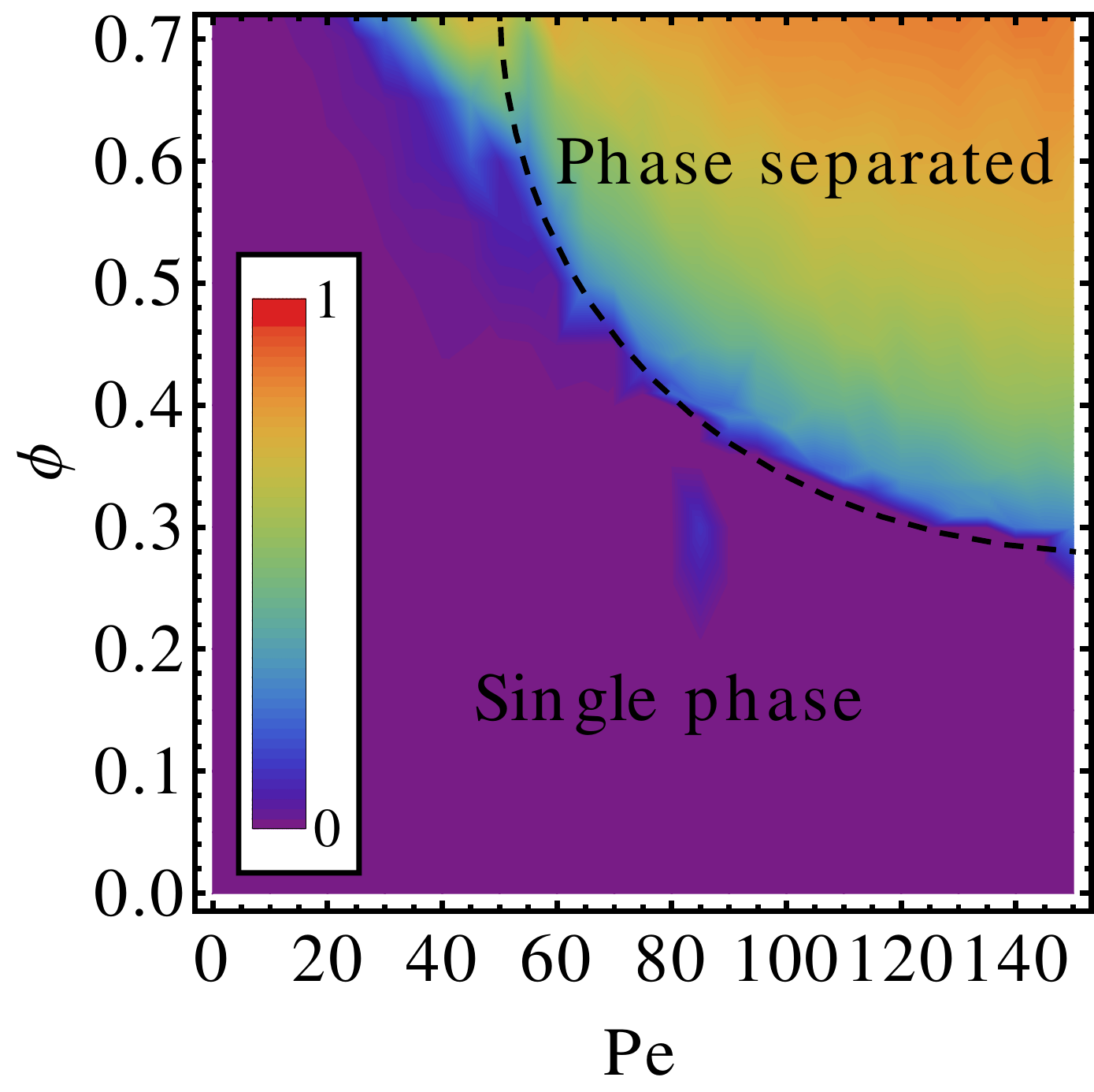}
  \includegraphics[width=.49\linewidth]{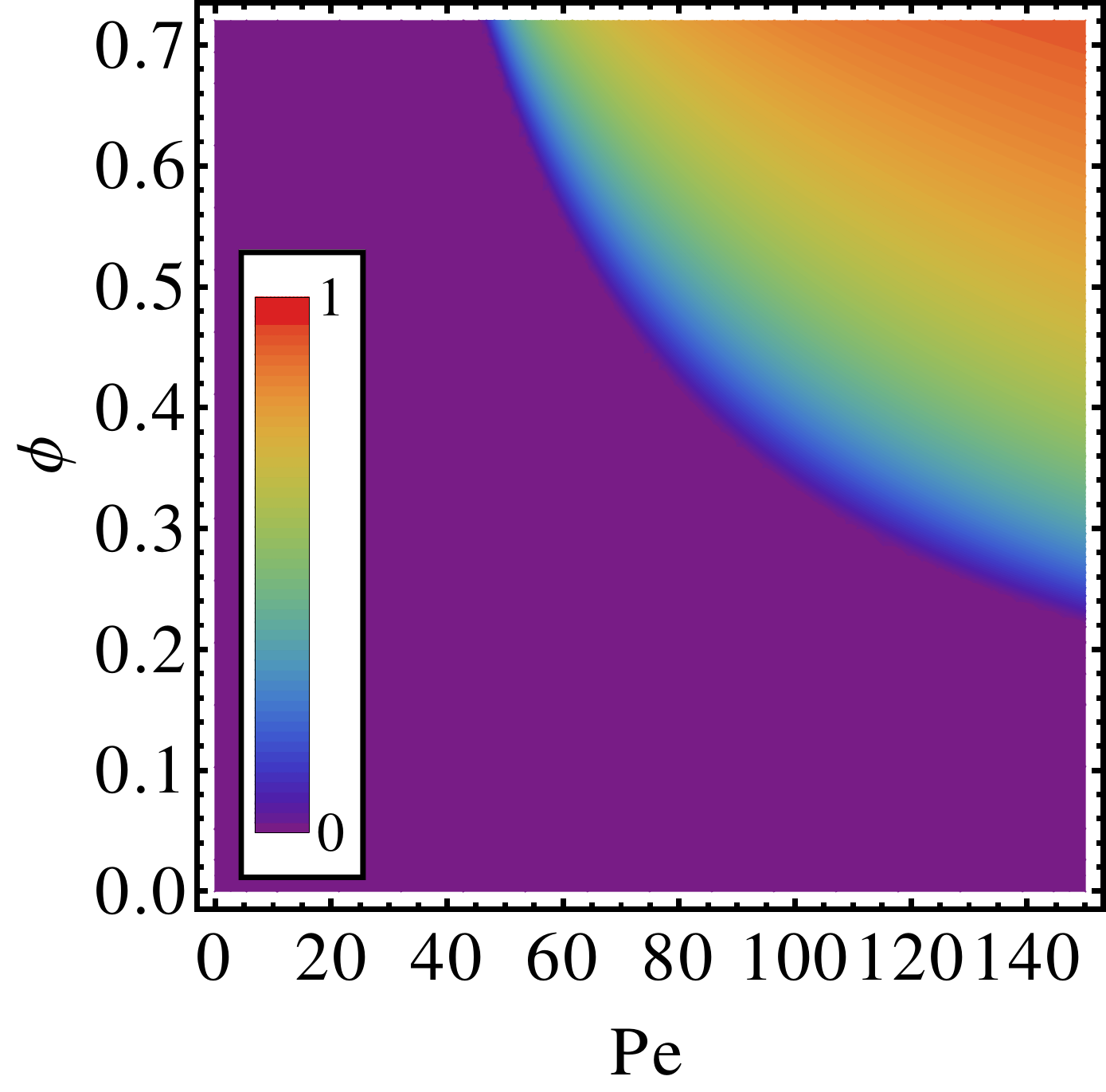}
  \caption{(color online) Left: Contour map of cluster fraction
    $\fc(\Pe, \phi)$ measured from simulations.  The dashed curve
    marks the approximate location of the binodal.  Right: Cluster
    fraction as predicted by our analytic theory
    (Eq. \ref{eq:cluster_fraction_theory}). These plots have been
    restricted to packing fractions that are low enough for the
    assumptions of our kinetic model to be valid, and for cluster
    identification to be unambiguous.}
  \label{fig:cluster_phase}
\end{figure}

\noindent \emph{The Phase-Separated Steady State}: To characterize the
steady state, we measured the fraction of particles in the dense phase
at time $100 \tau$ (Fig.~(\ref{fig:cluster_phase})).  In contrast with
recent work \cite{PhysRevLett.108.235702} which placed the phase
transition boundary at a constant density, we observe that this
cluster fraction is a nontrivial function of the system parameters
$\fc(\Pe, \phi)$. To understand this relationship we developed a
minimal model in which this function can be found analytically.  Let
us assume the steady state contains a macroscopic cluster which we
take to be close-packed. Particles in the cluster are stationary in
space but their $\theta_i$ continue to evolve diffusively. We treat
the gas as homogeneous and isotropic, and assume that a particle
colliding with the cluster surface is immediately absorbed.

Within this model, we can write the rate of absorption of particles of
orientation $\theta$ from the gas phase as $\kin(\theta) = \frac{1}{2
  \pi} \rhog \vp \cos \theta$, where $\rhog$ is the gas number
density.  Integrating yields the total incoming flux per unit length:
$\kin = \frac{\rhog \vp}{\pi}$. To estimate the rate of evaporation,
note that a particle on the cluster surface will remain there so long
as its self-propulsion direction remains ``below the horizon'', i.e.,
${\bf\hat{n}}\cdot\boldsymbol{\hat{\nu}}<0$, where ${\bf\hat{n}}$ is
normal to the surface. When its direction moves above the horizon, it
immediately escapes and joins the gas. This rate can be calculated by
solving the diffusion equation in angular space with absorbing
boundaries (for clusters large enough to treat the interface as flat,
at $\pm \frac{\pi}{2}$) and initial condition given by the
distribution of incident particles: $\partial_t P(\theta, t) = \Dr
\partial_\theta^2 P(\theta, t)$, with $P(\pm\frac{\pi}{2}, t) = 0$ and
$P(\theta, 0) = \frac{1}{2} \cos \theta$. Further, the departure of a
surface particle creates a hole through which subsurface particles
(whose $\boldsymbol{\hat{\nu}}_i$ may point outwards) can escape.
With $\kappa$ we denote the average total number of particles lost per
escape event, which we treat as a fitting parameter.  The total
outgoing rate is then $\kout = \frac{\kappa \Dr}{\sigma}$.

Equating $\kin$ and $\kout$ yields a steady-state condition for the
gas density: $\rhog = \frac{\pi \kappa \Dr}{\sigma \vp}$.  $\rhog$ can
be eliminated in favor of $\fc$, yielding (in terms of our
dimensionless parameters):
\begin{equation} \label{eq:cluster_fraction_theory}
  \fc = \frac{4 \phi \Pe - 3 \pi^2 \kappa}{4 \phi \Pe
    - 6 \sqrt{3} \pi \kappa \phi}
\end{equation}
This function is plotted in Fig.~(\ref{fig:cluster_phase}) with
$\kappa = 4.5$, in good accord with our simulation results.  Further,
the condition $\fc=0$ allows us to deduce a criterion for the onset of
clustering. Restoring dimensional quantities, this condition gives
$\phi \sigma \vp \sim \Dr$. Note that $\phi \sigma \vp$ is a collision
frequency; thus the system begins to cluster at parameters for which
the collision time becomes shorter than the rotational diffusion time.

The mechanism we have presented here is purely kinetic and requires
only an intuitive picture of local dynamics at the interface.  An
alternative view has been described by Tailleur and Cates
\cite{Tailleur2008,2012arXiv1206.1805C} who subsume all interactions
into a density-dependent propulsion velocity $v(\rho)$ which decreases
with density as collisions become more frequent.  From this they
construct an effective free energy which shows an instability in the
homogeneous phase if $v(\rho)$ falls quickly enough.  In a sense our
kinetic model represents an extreme case of this picture in which
$v(\rho)$ contains a step function such that free particles are
noninteracting, and particles in a cluster are completely trapped (see
Fig.~\ref*{Supplement-fig:d_eff} in \cite{SupplementalInformation}).

\begin{figure}[tbp]
  \raisebox{8pt}{\includegraphics[width=.445\columnwidth]{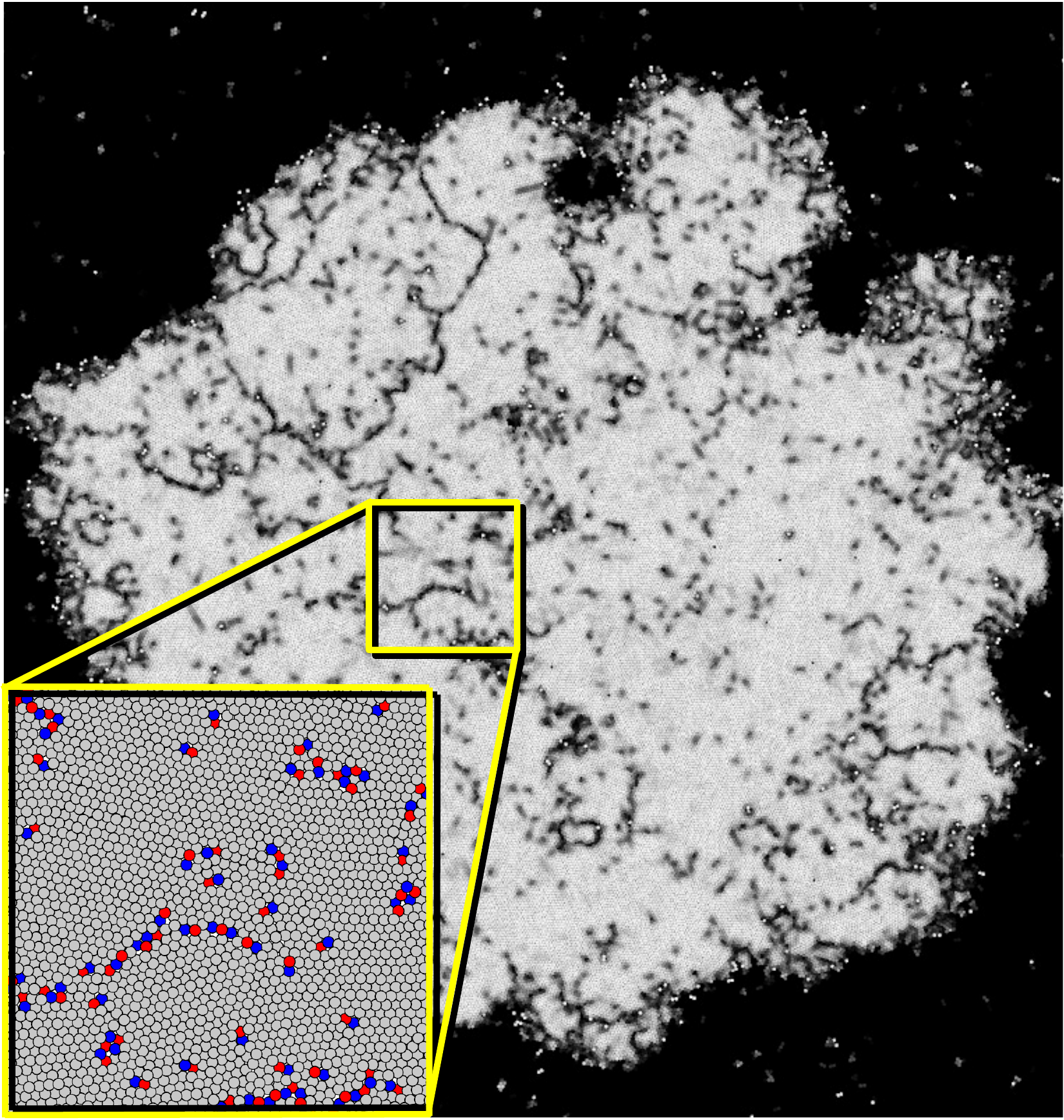}}
  \includegraphics[width=.54\columnwidth]{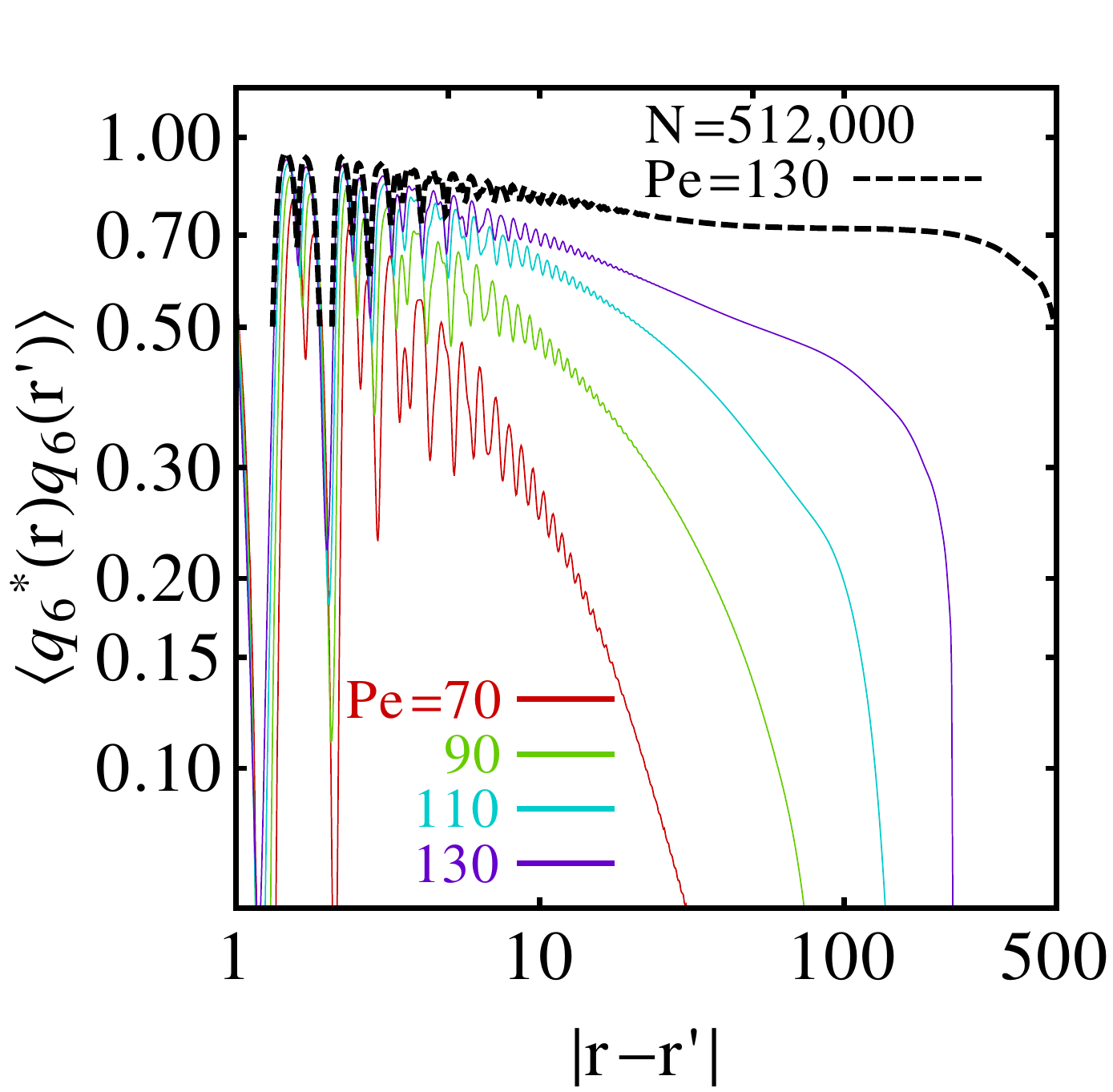}
  \caption{(color online) Left: Defect structures in a large cluster.
    Regions of high crystalline order (white) coexist with isolated
    and linear defects (dark).  The color of each particle indicates
    its $|\q6|$.  Inset shows pairs of 5/7 defects (red/blue).  Right:
    Log-log plot of the correlation function $\langle
    \q6^*(\mathbf{r}) \q6(\mathbf{r}')\rangle$ for clusters at various
    P\'eclet numbers in systems with $N=128{,}000$, showing a
    transition from liquid-like exponential to hexatic-like power-law
    decay as activity is increased.  For systems with $N=512{,}000$
    (black dashed line), a crystal-like plateau is also observed.}
  \label{fig:q6}
\end{figure}

\noindent \emph{Structure of the Dense Phase}: Since the system is
composed of monodisperse spheres, the dense phase is susceptible to
crystallization \cite{PhysRevLett.108.168301}.  As shown in
Fig.~(\ref{fig:summary}) the static structure factor of the cluster
interior shows a liquid-like isotropy at low $\Pe$, but develops
strong sixfold symmetry as activity is increased.  Further, the radial
distribution function shows clear peaks at the sites of a hexagonal
lattice (see Fig.~\ref*{Supplement-fig:radial_distribution} in
\cite{SupplementalInformation}) which sharpen and increase in number
as $\Pe$ is raised. We also measured the bond-orientational order
parameter $\q6(i) = \frac{1}{|\mathcal{N}(i)|} \sum_{j \in
  \mathcal{N}(i)} e^{i 6 \theta_{ij}}$, where $\mathcal{N}(i)$ runs
over the neighbors of particle $i$ (defined as being closer than a
threshold distance), and $\theta_{ij}$ is the angle between the
$i$-$j$ bond and an arbitrary axis (Fig.~(\ref{fig:q6})). We find a
structure characterized by large regions of high order with embedded
defects that are predominantly 5-7 pairs (Fig.~(\ref{fig:q6}a) inset
and \ref*{Supplement-fig:defects-movie} in
\cite{SupplementalInformation}).  Next, we examined the correlation
function $\left\langle \q6^*(\boldsymbol{r}) \q6(\boldsymbol{r}')
\right\rangle$ (Fig.~\ref{fig:q6}) which exhibits a liquid-like
exponential decay for systems of low activity, while at higher
activity the decay slows to a power law which is indicative of a
hexatic \cite{2002dgcm.bookN}.  A further transition to a crystal-like
plateau is observable in larger systems (see Fig. (\ref{fig:q6}) and
\ref*{Supplement-fig:q6correlation-size} and
\ref*{Supplement-fig:defect-density} in
\cite{SupplementalInformation}). In all cases, this material is unique
in that it is held together by active forces alone, and that the
arrest of motion is due to frustration. In this sense it is similar to
amorphous materials such as granular packs as reflected by the highly
heterogeneous stress distribution (Fig.~(\ref{fig:summary}))
\cite{doi:10.1021/jp809768y}.

\noindent \emph{Dynamics in the Dense Phase}: Within the active solid
material, self-propulsion forces continuously evolve by rotational
diffusion, breaking local force balance and leading to defect
formation and migration (see \ref*{Supplement-fig:defects-movie} in
\cite{SupplementalInformation}).  A compelling way to view the motion
produced by this athermal process is a simulated FRAP experiment
\cite{blaaderen:4591}, in which particles within a contiguous region
are tagged, making subsequent mingling of tagged and untagged
particles visible (see \ref*{Supplement-fig:frap-2.avi} in
\cite{SupplementalInformation}). To quantify this behavior, we
measured the mean square displacement (MSD) of particles in the
cluster interior.  As shown in Fig.~(\ref{fig:summary}), we observe
subdiffusive motion on short timescales, followed by a superdiffusive
regime, returning to diffusive motion on long timescales.  The
exponents of the subdiffusive and superdiffusive motion ($\frac{1}{2}$
and $\frac{3}{2}$, respectively) are well-conserved across a wide
range of propulsion strengths. Note that an isolated self-propelled
particle will exhibit diffusive, ballistic and diffusive behavior on
time scales $t < \frac{4 D}{\vp^2}$, $\frac{4 D}{\vp^2} < t <
\frac{1}{\Dr}$ and $t > \frac{1}{\Dr}$ respectively (see
Fig.~\ref*{Supplement-fig:msd} in
\cite{SupplementalInformation}). These dynamical regimes are modified
by the active solid environment; in particular, the ballistic regime
is modulated by ``sticking'' events as the particle is localized in
crystal domains, resulting in the observed L\'evy-flight-like behavior
\cite{klafter:33, doi:10.1080/00018737800101474}.

\begin{figure}[tbp]
  \includegraphics[width=0.495\linewidth]{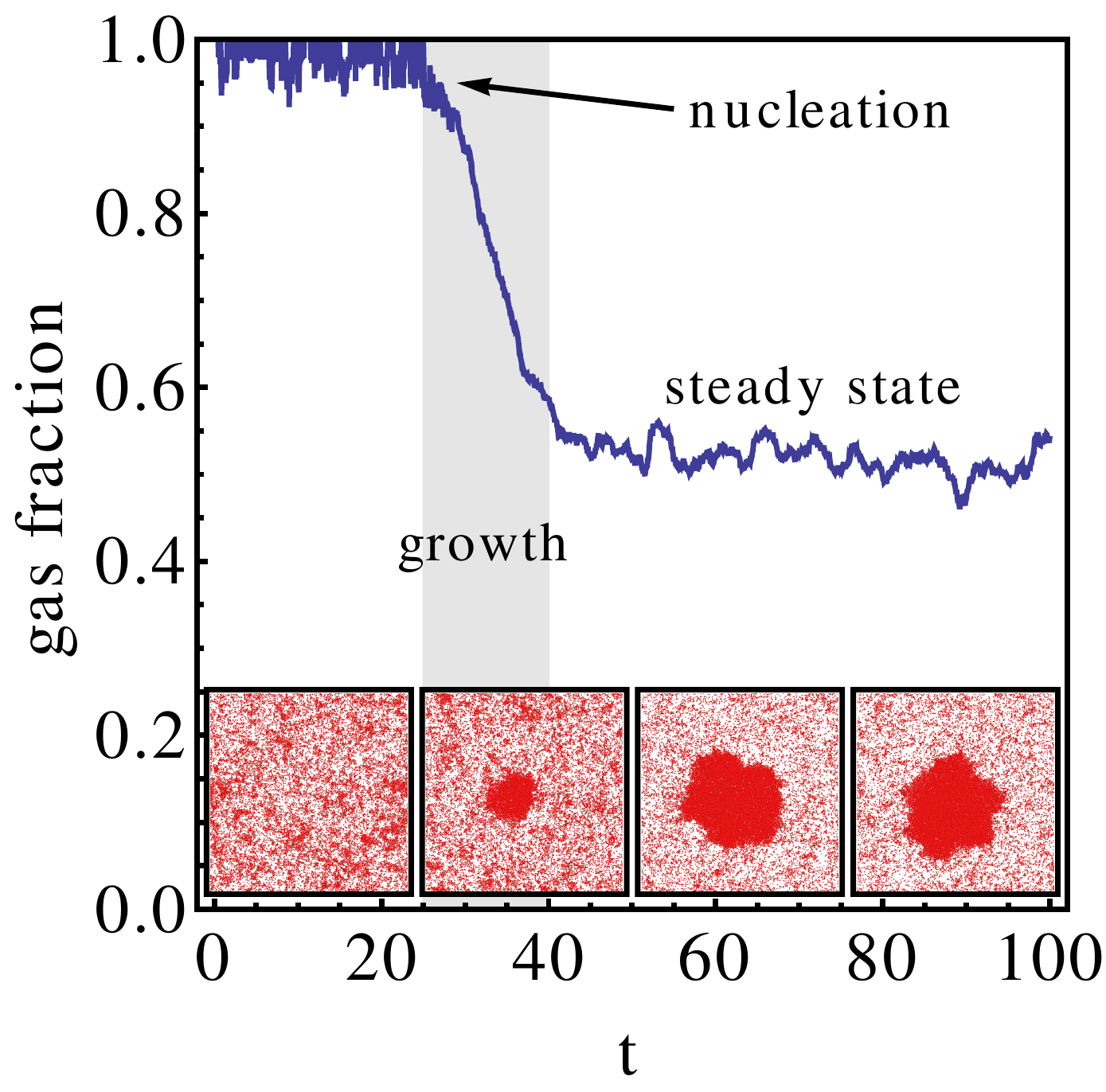}
  \includegraphics[width=0.485\linewidth]{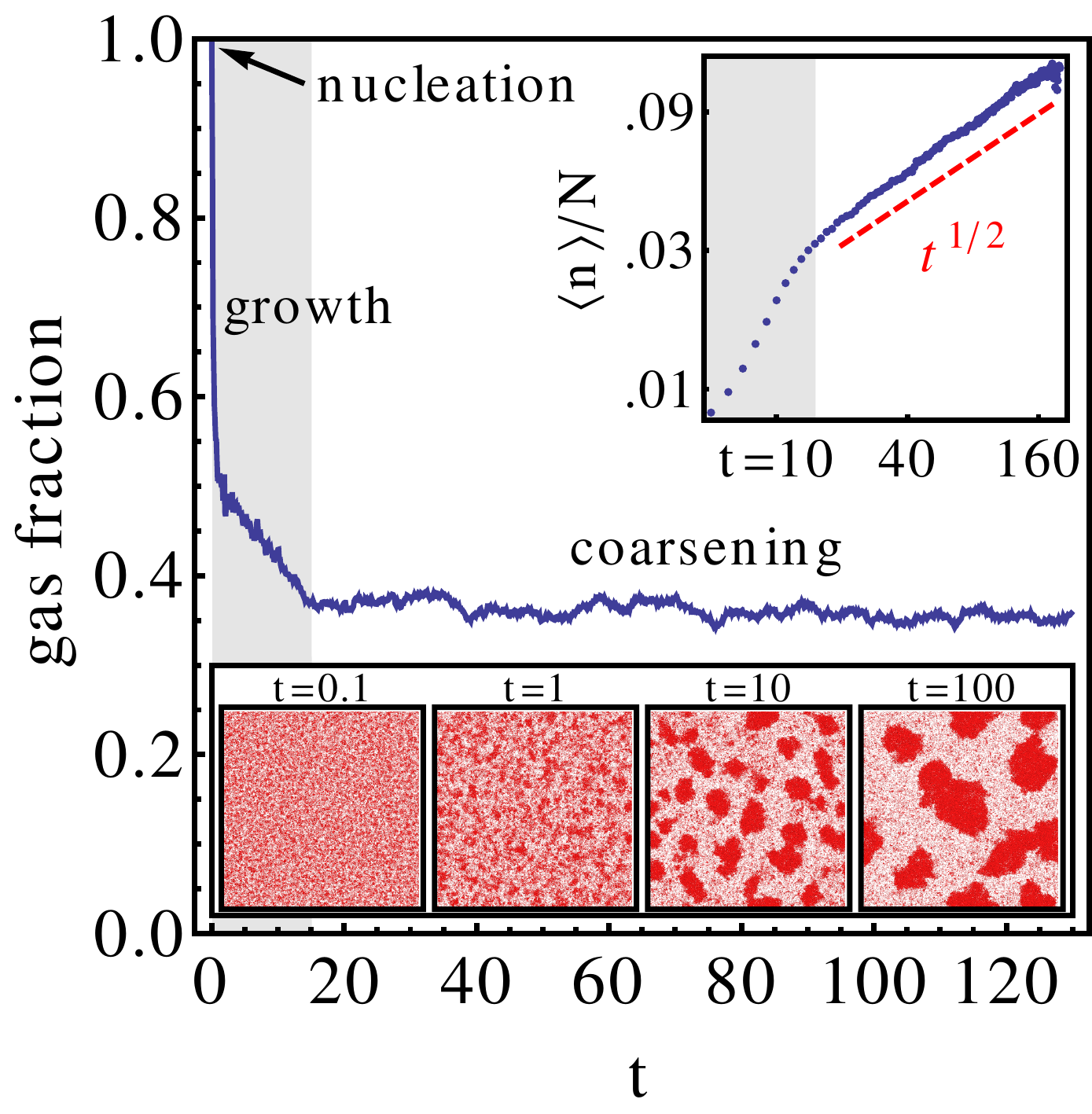}
  \caption{(color online) Examples of phase separation kinetics. Left:
    A system with $\Pe=100$, $\phi=0.45$ in which a delayed nucleation
    event leads quickly to steady-state.  For shallowly-quenched
    systems, the nucleation time can be long enough that artificial
    seeding is needed to make nucleation computationally accessible.
    Right: A system with $\Pe=80$, $\phi=0.6$ where spinodal
    decomposition leads to a coarsening regime which slowly evolves
    towards steady-state. Inset shows mean cluster size scaling
    approximately as $t^\frac{1}{2}$. (see
    \ref*{Supplement-fig:coarsening-movie} and
    \ref*{Supplement-fig:coarsening} in
    \cite{SupplementalInformation}).}
  \label{fig:nucleation_growth_coarsening}
\end{figure}

\noindent \emph{Kinetics of Phase Separation}: Despite the athermal
origins of phase separation in this system, simulations quenched to
parameters within the binodal experience familiar phase separation
kinetics (Fig.~(\ref{fig:nucleation_growth_coarsening})).  Systems
quenched close to the binodal exhibit a nucleation delay which can be
long enough that artificial seeding is necessary for phase separation
to be computationally accessible.  Systems quenched more deeply
undergo spinodal decomposition, leading to a coarsening regime in
which the mean cluster size scales surprisingly as $t^\frac{1}{2}$,
with a corresponding length scale $\mathcal{L}(t) \sim t^\frac{1}{4}$
(Fig.~(\ref{fig:nucleation_growth_coarsening}) inset, also see
\ref*{Supplement-fig:coarsening} in
\cite{SupplementalInformation}). This differs from the standard 2D
coarsening exponents, but matches recent simulation results for the
Vicsek model and related active systems \cite{PhysRevLett.108.238001}.
This result should be viewed as preliminary due to the
limited range of our data, but nevertheless this unexpected similarity
between the coarsening of point-particles with polar alignment and
that of spheres with no alignment suggests a deep relationship between
these very different types of systems.  Future work is needed to
uncover the origins of these scaling exponents and their implications
for universality in active fluids.

\noindent \emph{Summary}: A fluid of self-propelled colloidal spheres
exhibits the athermal phase separation that is intrinsic to active
fluids and is a primary mechanism leading to emergent structures in
diverse systems \cite{Gopinath2011,Cates2010}. We have shown that the
physics underlying this phase behavior can be understood in terms of
microscopic parameters.  From a practical perspective, our simulations
show that the active solid dense phase exhibits a combination of
structural and transport properties not achievable in a traditional
passive material. Further development of experimental realizations of
this system (e.g. Ref. \cite{2012arXiv1202.6264T}) will advance the
development of materials whose phase behavior, rheology, and transport
properties can be precisely controlled by activity level.

{\bf Acknowledgments:} This work was supported by NSF-MRSEC-0820492
(GSR, MFH, AB), as well as NSF-DMR-1149266 and NSF-1066293 and the
hospitality of the Aspen Center for Physics (AB). Computational
support was provided by the Brandeis HPC.

\bibliographystyle{apsrev4-1}
\bibliography{paper}

\end{document}